\documentclass[twocolumn]{aastex631}

\usepackage{amsmath}

\usepackage{rotating}
\usepackage[flushleft]{threeparttable}

\begin{document}

\title{The Red Supergiant Problem: As Seen from the Local Group's Red Supergiant Populations}

\author{Sarah Healy}
\affiliation{Center for Neutrino Physics, Department of Physics, Virginia Tech, Blacksburg, VA 24061, USA}

\author{Shunsaku Horiuchi}
\affiliation{Center for Neutrino Physics, Department of Physics, Virginia Tech, Blacksburg, VA 24061, USA}
\affiliation{Kavli IPMU (WPI), UTIAS, The University of Tokyo, Kashiwa, Chiba 277-8583, Japan}

\author{Chris Ashall}
\affiliation{Center for Neutrino Physics, Department of Physics, Virginia Tech, Blacksburg, VA 24061, USA}
\affiliation{Institute for Astronomy, University of Hawai'i at Manoa, 2680 Woodlawn Dr., Hawai'i, HI 96822, USA }



\begin{abstract}
 The red supergiant (RSG) problem, which describes the apparent lack of high-luminosity progenitors detected in Type II supernova (SN) pre-images, has been a contentious topic for two decades. We re-assess this problem using a new RSG population of the Milky Way supplemented with RSGs from other galaxies in the Local Group. In particular, we quantify the uncertainties inherent to assumptions made regarding the star’s temperature or spectral type and the corresponding bolometric correction. We find that only M3 or later RSGs reproduce the steepness seen from the SN II pre-imaged sample.  To assess the significance of the RSG problem, we build a metallicity-weighted cumulative luminosity distribution of M3 or later RSGs and directly compare it to the luminosity distribution of SN II pre-imaged progenitors. We find no evidence of missing high-luminosity pre-imaged progenitors since the uncertainties on the pre-imaged SN progenitors and single-band derived luminosity are too large to meaningfully infer population differences.  
\end{abstract}


\section{Introduction} \label{sec:intro}

With the introduction of large-scale transient surveys, hundreds of core-collapse supernovae (CCSNe) are discovered each year. Even though most occur in distant galaxies where only indirect probes of the core-collapse (CC) process are possible, enough have been discovered in neighboring galaxies and with pre-explosion imaging, enabling the identification of RSGs as the direct progenitors of Type IIP SNe \citep{smith_2009,smith_2014,smith_2011,maund_2004}. Thus, these transient sources are the terminal explosions of RSGs which started out with initial mass more massive than $\approx$ 8 $\rm M_\odot$, assuming single star evolution, that undergo nuclear burning until the iron peak leads to the unbinding of their outer layers via CC and neutrino driven explosions \citep{burrows_1995}. 

However, archival imaged detections of Type IIP progenitors find only objects with log($\rm L/L_\odot) <$ 5.24 in contrast to the observed general population of RSGs whose luminosity extends to log($\rm L/L_\odot) \approx$ 5.6. Early discussions \citep{Kochanek:2008mp} eventually led to the discrepancy being coined the RSG problem \citep{smartt_2009}. Over time, the number of detected SN progenitors has increased, and confirmations of the disappearance of the SN progenitor candidates have confirmed them as SN progenitors \citep[e.g.,][]{maund_2015}. Based on number statistics, the RSG problem has been reported to be up to several sigma significance \citep{smartt_2009}.

\begin{deluxetable*}{lccccccccccc} 
\tabletypesize{\scriptsize}
\tablewidth{\textwidth} 
\tablenum{1}
\tablecaption{Characteristics of RSG candidates in the Local Group
\label{Tab:Tcr}
} 
\tablehead{ \colhead{Galaxy} & \colhead{Metallicity} & \multicolumn{3}{c}{$\rm Teff$ [K]$^{a}$ } & \multicolumn{3}{c}{Sample Size}  & \multicolumn{3}{c}{$\rm Log(\frac{L_{max}}{L_\odot})$} & \colhead{Ref}\\
 \colhead{} & \colhead{[$Z_\odot$]} &
  \colhead{K0+} &  \colhead{M0+} & \colhead{M3+} &  
  \colhead{K0+} &  \colhead{M0+} & \colhead{M3+} &  \colhead{K0+} &  \colhead{M0+} & \colhead{M3+} 
 &  \colhead{}}
\colnumbers
\startdata 
    M31 & 1.5-2 & 4185 & 3850 & 3625 & \begin{tabular}{c} 4239 \\ 5587  \end{tabular} & \begin{tabular}{c} 556 \\ 628  \end{tabular} & \begin{tabular}{c} 200 \\ 235 \end{tabular} &  
    \begin{tabular}{c} 5.75 \\ 5.46 \end{tabular}  & \begin{tabular}{c} 5.75 \\ 5.36 \end{tabular} & 
    \begin{tabular}{c} 5.69 \\ 5.36 \end{tabular} &
    \begin{tabular}{c} R21\\ M23 \end{tabular}\\
    MW & 1 & 4185 & 3750 & 3605 & 461 & 364 & 117 &   $6.02^{+0.2}_{-0.18}$ & $5.64^{+0.41}_{-0.31}$ & $5.64^{+0.41}_{-0.31}$  &\textbf{This Work}\\
    M33 & 0.9  &  4185$^{b}$ & 3750 & 3605 &  \begin{tabular}{c} 2815 \\2457 \end{tabular} &   \begin{tabular}{c} 711 \\ 1052 \end{tabular} & \begin{tabular}{c} 91 \\161 \end{tabular} & 
    \begin{tabular}{c} 5.47 \\ 5.52 \end{tabular} &  \begin{tabular}{c} 5.47 \\ 5.52 \end{tabular} &  \begin{tabular}{c} 5.47 \\ 5.52 \end{tabular} & 
    \begin{tabular}{c} R21\\ M23 \end{tabular}\\
    LMC & 0.4   &  4550  & 3950 & 3545
    & \begin{tabular}{c} 2273 \\ 4075 \end{tabular} &  \begin{tabular}{c} 438 \\ 1150 \end{tabular} &  \begin{tabular}{c} 61 \\ 155 \end{tabular} & 
     \begin{tabular}{c} 5.54 \\ 5.59 \end{tabular} & \begin{tabular}{c} 5.54 \\ 5.59 \end{tabular} & \begin{tabular}{c} 5.53 \\ 5.48 \end{tabular} &
    \begin{tabular}{c} Y21 \\ M23 \end{tabular}\\
    SMC & 0.15  &     4372 & 3850  & 3325$^{c}$ & \begin{tabular}{c} 1506 \\ 1553 \end{tabular} & \begin{tabular}{c} 139  \\ 22 \end{tabular} & \begin{tabular}{c} 6 \\ 3 \end{tabular} &
    \begin{tabular}{c} 5.54 \\ 5.49 \end{tabular} &  \begin{tabular}{c} 5.32 \\ 5.14 \end{tabular} & \begin{tabular}{c} 5.02 \\ 4.91 \end{tabular} & 
    \begin{tabular}{c} Y23 \\ M23
    \end{tabular}
\enddata
\tablecomments{Summary of the Local Group galaxies analyzed and the various effective temperature scales derived for each (3-5), the size of the samples based of spectral type cuts as described in Sec \ref{mw list} (6-8), and the maximum luminosity of each sample (9-11). The references are shown in col (12) as M23: \cite{Massey_2023}, Y21: \citep{yang_2021lmc}, R21: \cite{Ren_2021m33m31} Y23: \cite{yang_2023}. }
\footnotesize{$^a$ Effective temperature Scales built from \cite{Levesque_2005,Massey_2009,Levesque_2006} with average uncertainty $\rm\sim 100$K.}

\footnotesize{$^b$ effective temperature scale from Milky Way \cite{Levesque_2005} used in absence of scale for M33}

\footnotesize{$^c$ effective temperature scale for SMC incomplete so $\rm T_{\rm eff}$ assigned to M2 is used instead}
\end{deluxetable*}

\vspace{-9mm}

Various solutions to the RSG problem have been proposed, e.g., that failed SNe---CC which proceeds to black hole formation without a luminous CCSN---provide a new channel for high-mass stars, short-lived phases of extreme mass loss rates \citep{smith_2009,smith_2014}, or binary interaction stripping the RSGs envelopes \citep{podsiadlowski_1992,eldridge_2013,Zapartas_2017}, and the implications for CCSN theory \citep{Horiuchi:2014ska}. In parallel, research into the significance of the RSG problem has also been made. For example, deriving the parameters of the progenitor from pre-explosion archival imaging requires a large number of assumptions about spectral type and circumstellar dust and introduces uncertainties in telescope sensitivity \citep{Strotjohann_2024}, fluctuation due to measurement uncertainties \citep{davies_2020a_rsg_p_ii} and bolometric corrections either in the optical \citep{beasor_2024} or infrared (IR) \citep{davies_beasor_2018_bc} for which various efforts have been made to quantify their effects on the resulting luminosity.

Despite these efforts, challenges remain. For example, the sample size of RGSs being compared remains limited to $\lesssim 300$, in contrast to an expected size on the order of a few thousand per galaxy. The reason for this is the limited availability of well-sampled stellar energy distributions (SEDs). However, it raises concerns of biases. Another challenge is that the majority of SN pre-images have only single-band imaging. This is a concern in particular for RSGs since a significant portion of the RSG characteristics are missing, making accurate estimates of the stellar properties uncertain. Finally, metallicity differences must be appropriately addressed. As the role of metallicity on the luminosity of RSGs is still debated, the metallicity of the pre-imaged SN progenitors should be matched by the sample of RSGs before a proper comparison can be made.

In this paper, we use a well-mapped sample of RSGs from the Local Group which covers a range of metallicities, to reassess the RSG problem. This enables us to provide an improved look at RSGs before CC and allows for direct comparison to detected progenitors and across the entire range of their luminosity function. To these ends, we collect large samples of RSGs, re-quantify uncertainties due to optical and near-infrared (NIR) single-band derived luminosity, and constrain the spectral types characteristic of CCSNe progenitors.

The paper is organized as follows. In Section \ref{sec: pro samples}, we discuss the physical parameters, mainly luminosity and effective temperatures, of various samples of RSGs and pre-explosion imaged RSGs. In Section \ref{sec: luminosity functions}, we discuss the effect metallicity has on the RSG luminosity functions and make direct comparisons between pre-explosion RSG, the general observed population of RSGs, and bias-corrected cumulative distribution functions. We discuss our results and their implications for the RSG problem in Section \ref{sec: results and dicussion} and conclude in Section \ref{sec: conclude}.

\begin{figure*} 
 \centering
 \includegraphics[width=0.9\linewidth]{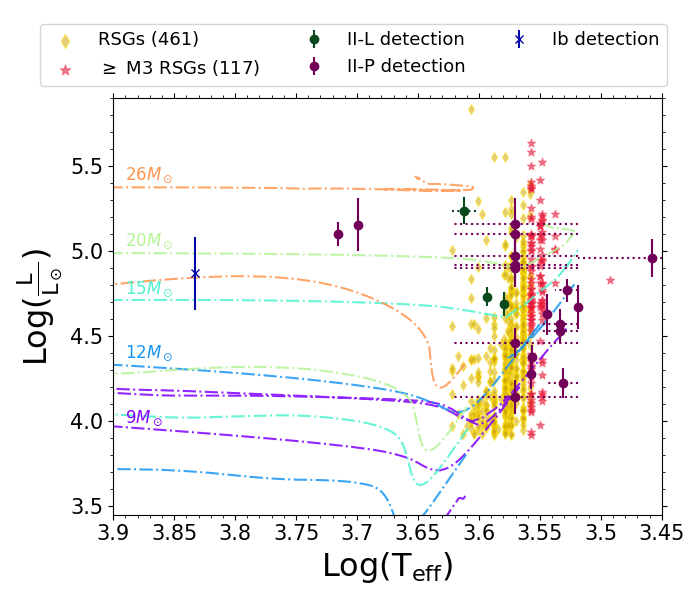}
 \caption{Hertzsprung-Russell diagram focused on the RSG branch region. Confirmed progenitors of Types II-L, II-P, and Ib SNe are shown in green, maroon, and blue, respectively. Plotted in the background as yellow diamonds and red stars are the location of \cite{healy_2023}'s Galactic RSG sample of spectral type K0 or later and M0 or later, respectively. Dash-dotted lines are MIST single star evolutionary tracks of \cite{Choi_2016}.}\label{fig: hr}
\end{figure*}

\section{Progenitor Samples} \label{sec: pro samples}

Previous attempts to investigate the RSG problem focused on the pre-imaged SN progenitors and comparison to a limited sample of RSGs. In order to account for larger possible ranges of RSGs properties, in particular as they evolve, we begin by compiling complete samples of observed RSGs within 0.850 Mpc and all SN pre-imaged progenitors out to $\rm \sim$30 Mpc. We utilize values from the literature or estimate sample characteristics, like luminosities and effective temperature, allowing for comparison and consistency tests detailed in later sections. Below, we discuss the various datasets we use.

\subsection{Milky Way RSGs} \label{mw list}

Large-scale surveys in the past decades have expanded the number of known massive stars, particularly RSGs, and allowed for a better understanding of the luminosity distribution of various populations (see Table \ref{Tab:Tcr}). While Local Group members like the Magellanic Clouds (MCs), M31, and M33 have the best sample in terms of completeness of RSGs, the metallicity of the Milky Way (MW) and its size provide an environment more representative of those of the detected pre-images of Type II SNe, which are biased towards more massive galaxies close to solar metallicity. 

Complete studies of the MW have so far been prevented by limitations in the detailed mapping of Galactic dust and its effect on observation. However, recent efforts to prepare for the next galactic supernova, likely the first chance to observe EM waves, neutrinos, and gravitational waves from a single event, has resulted in more extensive lists of Galactic RSG candidates designed for multi-messenger efforts \citep{healy_2023} presenting a statistically significant sample with which to reassess how well-detected progenitors match observed RSGs.

Previous efforts have looked into smaller Galactic RSG samples like \cite{Levesque_2005}'s MW sample with $\sim$80 optically identified RSGs \citep{davies_2018a_rsg_p}. However, by surveying the literature and using Gaia data, \cite{healy_2023} compiled a catalog of 578 MW RSGs, the largest catalog of its kind in the literature. Bolometric luminosity was derived from $\rm K_s$ band from the Two Micron All Sky Survey (2MASS) , and dust extinction was estimated using {\tt mwdust},
a 3D dust map \citep{bovy_2015}. The $\rm K_s$ was used to derive luminosity \citep{Skrutskie_2006} and was chosen for consistency and completeness discussed further in Section \ref{sec:NIR}, and the effective temperature of all RSG candidates were pulled to determine the luminosity function of the sample. Contaminants were filtered out with a combination of comparisons to stellar evolutionary tracks and observations of Galactic Asymptotic Giant Branch (AGBs) stars. Nevertheless, there are still likely AGBs within the sample. These were retained to maintain as many RSGs as possible to best fulfill the paper's goal of preparing for the next Galactic SN. For our purposes, this provides a sample of RSGs that is large but incomplete below $\rm log(L/L_\odot) =4.6$. This sample is labeled as ``this work''. In Fig.~\ref{fig: hr}, the sample is shown separated by spectral types K or early type RSGs and M or late type RSGs, referred to as K0+ depicted by 461 yellow diamonds and M3+ shown as 117 red stars, respectively.

\subsection{Local Group RSGs}

To avoid the issues of uncertain distances and high interstellar reddening that hindered the MW's completion, other members of the Local Group have been strategic targets for collecting RSG samples. Also, by combining them with the MW, more complete RSG samples can be obtained spanning metallicity to replicate the metallicity profile of the pre-imaged SN progenitors. This provides us with another metric to compare to that of SN pre-images and interpret the significance of the RSG problem.

Recent evidence \citep{higgins_2020_hd_lim,davies_2018b_lmc_hd_limit,Mcdonald_2022,gilkis_2021_hd} suggests that the Humphreys-Davidson limit \citep{h_d_limit_1979}, an empirical upper limit to massive star luminosity, and RSG luminosity functions are not metallicity dependent in contradiction to stellar evolution models. Without a more complete understanding, it is difficult to reliably predict how metallicity will affect each luminosity function. Thus, we pursue a data-driven comparison. 

We searched the literature finding two sets of RSGs from the MCs, M31, and M33 galaxies built using diverging methods, resulting in two versions with crucial differences in what portion of the RSG population is represented.  

We pull SMC RSGs from \cite{yang_2023}, LMC from \cite{yang_2021lmc} and M31 and M33 from \cite{Ren_2021m33m31}. The data is built from near-infrared photometric data from UKIRT/WFCAM \citep{irwin_2013} and the 2MASS point source catalog \citep{Skrutskie_2006} and use formulas from \cite{Neugent_2020_EQ} to determine effective temperatures and bolometric corrections. Newer versions are available but rely on SEDs with strict photometric quality criteria, removing enough objects to no longer classify as complete samples. An effort was made to retain faint and red objects, even highly reddened objects, resulting in colour cuts to remove foreground stars bluer and AGBs fainter than expected boundaries. The M31 and M33 samples should be complete as the lower limit of the RSG brightness in M31 and M33 is higher than the observationally complete magnitude, and those of the LMC and SMC are considered to be complete as well. We call this sample of 13,277 stars as the Ren list.

Additionally, \cite{Massey_2023} (and references therein) used near-IR photometry to identify RSGs and best fits to MARCS stellar atmosphere models to derive their effective temperatures and bolometric luminosities. As an update on previous work, a constant extinction based on spectral fitting by \cite{Massey_2009} per host galaxy was used, though it likely resulted in dusty RSGs' luminosities being underestimated. While the samples should be complete to a luminosity limit of $\rm log(L/L_\odot) =4.0$, strict colour cuts left less contamination from AGBs but removed RSGs with high reddening. We refer to this sample of 14,216 stars as the Massey list.

\subsection{Near-Infrared Derived Luminosity} \label{sec:NIR}

\begin{figure} 
 \centering
 \includegraphics[width=0.98\linewidth]{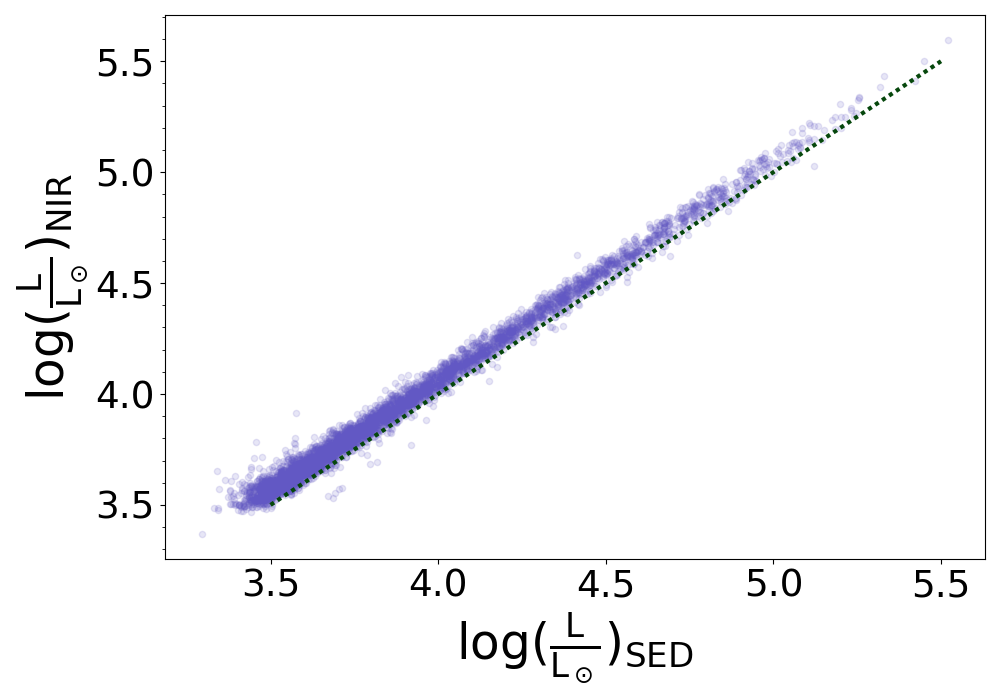}
 \caption{Luminosity derived from NIR photometry using method from \cite{neugent_2020a} versus from observed SED from \cite{Wen_2024}.}\label{fig: nir_lum}
\end{figure}

\begin{figure*}
    \centerline{\includegraphics[width=\textwidth]{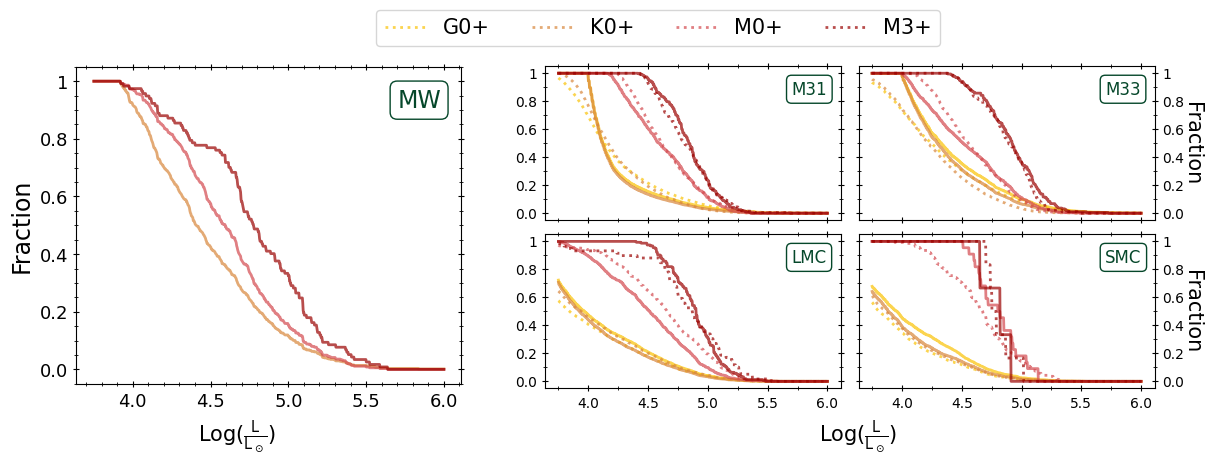}}
    \caption{Cumulative distribution functions for RSGs in the MW, M31, M33, LMC, and SMC, showing its progression as the sample restricted: groups G0+, K0+, and M3+ include all stars with spectral types later than $\geq$G0, $\geq$K0, and $\geq$M3, respectively. For each galaxy two independent samples are shown:  \cite{yang_2021lmc,Ren_2021m33m31,yang_2023}, (dotted) and \cite{Massey_2023} (solid).}\label{cum k0, m0, m3}
\end{figure*}

This work and the Massey list have luminosity already estimated based on NIR photometry, but the Ren sample was missing estimation of physical parameters. For consistency and completeness, though luminosities from SEDs are available in some cases, we derived luminosity for the Ren sample from NIR, mainly 2MASS J and $\rm K_s$, photometry. Compared to luminosity derived from SEDs, NIR overestimates but stays within $\rm \sim 0.1$ dex, as shown in Fig.~\ref{fig: nir_lum}. NIR also limits the effect of variability, which is $\rm \sim 0.1$ mag, but in optical, it can be as much as one mag. For effective temperature and bolometric corrections, we pull from the literature equations built specifically for each galaxy (M31 \& M33: \cite{Massey_2023},  LMC \& SMC: \cite{de_wit_2023}).

While primarily a RSGs sample, the combined list does contain some YSGs, so we split our samples into four sets: stars with effective temperatures later than or equal to G0, K0 or all RSGs, M0, and M3 or late-type RSGs (labeled as G0+, K0+, M0+ and M3+ respectively, as shown in Fig.~\ref{cum k0, m0, m3}). The difference in colour cuts, extinction estimates, and host galaxy environments produce varying maximum luminosities and cumulative luminosity distributions (CLDs) between samples. Table \ref{Tab:Tcr} shows our MW sample and the local group galaxies' characteristics, including maximum observed luminosity, average metallicity, and sample size. We also include our effective temperature cut-offs, in columns 3 (K0+), 4 (M0+) and 5 (M3+), for spectral types at each metallicity based on a combination of effective temperature scales \citep{Levesque_2005,Massey_2009,Levesque_2006}. In the absence of well-defined scales for M33, we use MW's as it has a similar average metallicity. The entire samples are within G0+, but the highest temperature varies between samples, so no specific limits are provided in Table \ref{Tab:Tcr}.

In Fig.~\ref{cum k0, m0, m3}, it can be seen that the CLDs of the populations from G0+ and K0+ have similar shapes, while those of M0+ and especially M3+ have steeper slopes and are shifted towards high luminosities. It is possible that this shift could be attributed to a missing number of fainter redder stars, which are more likely to be missed; however, this would imply missing objects in samples determined to be statistically or approximately complete. While this cannot be definitively ruled out, as observational completeness inherently carries some uncertainty, even in the worst case, this number is unlikely to be more than a few objects based on the criteria and methods used to build these samples. This, together with the uniformity of the increase in steepness across all metallicities and between the source lists, we do not believe the chance of missing low luminosity RSG has a significant influence on the resulting CLDs.

\subsection{RSGs Detected in SN Pre-imaging} \label{pre-imaging data}
Detected SN progenitors of II-P/L and II with measurements or meaningful limits were compiled from the literature, similar to \cite{Strotjohann_2024}. To this, we add SN2024ggi and update values for SN2023ixf since CSM dust estimates were revised \citep{neustadt_2024,xiang_2024,van_dyk_2024}. We add on estimates of metallicities either from nearby HII regions or their host galaxy's metallicity estimates in Table \ref{A.1}. We also include progenitors' effective temperatures either from the literature or based on assigned spectral type or range and RSG effective temperature scales for its associated metallicity. Along with calculated bolometric luminosities from the original single-band observations, bolometric corrections from \cite{davies_beasor_2018_bc}, distance modulus when available, and similarly to \cite{Strotjohann_2024} we estimated a limiting luminosity to see the sensitivity of each detection. Limits are either directly from limited magnitudes quoted from the literature or assuming the error on the observation is dominated by the noise level of the background and ignoring additional potential uncertainties, so the errors can be converted to $\rm 3\sigma$ limiting magnitude, shown in Table \ref{A.2}.

The estimated effective temperature and luminosity based on pre-explosion imaging of the CCSN in our detected progenitor sample are shown in Fig.~\ref{fig: hr}, shown with MIST single-star evolutionary tracks at solar metallicity \citep{Choi_2016} and MW RSG candidates, detailed in Section \ref{mw list}, split into K0+ (yellow diamonds) and M3+ (red stars). Unless multi-band imaging is available, assuming effective temperature and luminosity are highly uncertain, as can be seen from the large error bars of progenitors in Fig.~\ref{fig: hr} and the number of objects classified no more precisely than as ``RSG" in \ref{A.2}. Note there is a region around $\rm \gtrsim 5.4$ where there are no detected SN progenitors, yet there are observed MW RSG candidates: this is the RSG problem (see also Section \ref{mw list}).

\subsection{Optical band derived Luminosity} \label{sec:I lum}

\begin{figure} 
 \centering
 \includegraphics[width=0.98\linewidth]{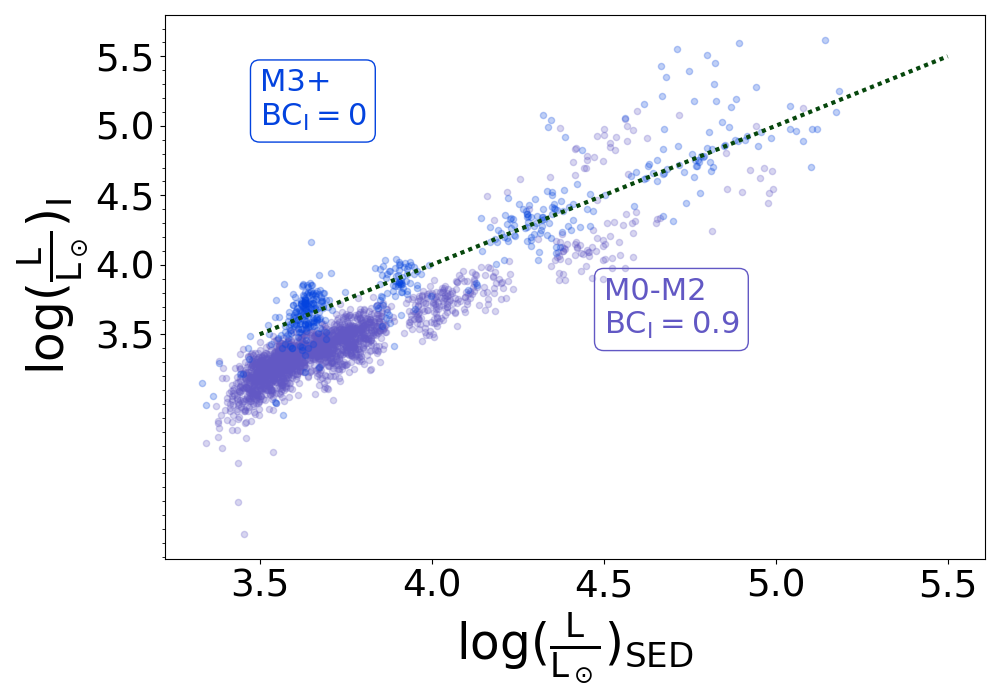}
 \caption{Luminosity derived from optical I band photometry versus from observed SED from \cite{Wen_2024}. The sample is broken into two groups based on the available bolometric corrections denoted in the square boxes and colored to match the associated points. M3+ is shown in blue, and M0-M2 is shown in purple. }\label{fig: I_lum}
\end{figure}

Optical, specifically the F814W band, photometry is the most often pre-explosion imaging found for detecting SN progenitors. In general, optical wavelengths contain unreliable information for RSGs as they are bluewards of the RSG's peak emission and any re-emitted flux due to the surrounding dust. Using M31 139 RSGs with well-observed optical to mid-IR photometry, \cite{beasor_2024} find that single-band photometry systematically underestimates the luminosity of RSG by an average of 0.3 dex when M0 is assumed for the entire sample and 0.08 dex similarly for M5 with an average dispersion on the order of 0.2 dex.

We repeat this study with a larger sample of 2,536 M0+ RSGs from the LMC, pulling I-band ($\rm \approx F814W$) photometry from the Magellanic Clouds Photometric Survey (MCPS) \citep{Zaritsky_2004}. Extinction came from the dust map of \cite{chen_2022_10.1093/mnras/stac072}, an extinction coefficient from the classical CCM89 model \citep{Cardelli_1989ApJ...345..245C}, and $\rm R_V = 3.41$ \citep{Gordon_2003}. As bolometric corrections for the I-band are fractured with values determined for M0 and M5+ but no singular equation that covers the entire range of spectral types, we break the sample into groups of M0-M2 and M3+. For M0-M2, we use $\rm BC_I=0.9$ \citep{smartt_2009,smartt_2015}, and the luminosity derived is represented by the purple points in Fig.~\ref{fig: I_lum}. For our M3+ sample, we use $\rm BC_I=0$ \citep{davies_beasor_2018_bc}, the M5+ bolometric correction built to represent the late-type RSGs. The resulting luminosity ($\rm log(L/L_\odot)_I$) estimates are shown by the blue dots in Fig.~\ref{fig: I_lum}. It is important to note that different treatments between our sample and \cite{beasor_2024}. The spectral types are known for these RSGs, so bolometric corrections are specifically applied to RSGs with the associated spectral types rather than an entire sample treated as a singular spectral type with a singular bolometric correction.

In either case, the $\rm Lum_I$ has a wide dispersion, significantly wider than that of the $\rm log(L/L_\odot)_{NIR}$ in Fig.~\ref{fig: nir_lum}. For M0-M2, $\rm log(L/L_\odot)_I$ is underestimated in comparison to the luminosity from the observed SEDs up until $\rm log(L/L_\odot)_{SED} = 4.25 $ where the average is much closer to the 1-to-1 line, but the dispersion has widened to $\rm \sim0.5$ dex. However, the M3+ estimates show a widening dispersion and a slightly increased deviation above the 1-to-1 line as luminosity increases.

\section{RSG Luminosity Distributions} \label{sec: luminosity functions}

While it is hard to draw information from comparing individual RSG and SNe, studying the complete samples of RSGs in terms of population behaviors can help resolve the issues with the missing high-mass pre-imaged SN progenitors. Mass estimates for our samples are unreliable, especially due to the derivation of luminosity from single-band photometry, but effective temperature and luminosity distributions allow for comparison between RSG and pre-imaged SN progenitor populations. First, we look into how observed RSGs luminosity distribution changes as the effective temperature lowers and the stars evolve, then into those pre-imaged SN progenitors, and determine the effect metallicity has on both.

\cite{davies_2018a_rsg_p} showed RSGs likely evolve to later spectral types as they approach SNe and suggested SN progenitors in the absence of other knowledge should be assumed to be late types. In order to determine the spectral characteristic of Type II progenitors at their deaths, we compare to subpopulations of RSGs with varying $\rm T_{\rm eff}$ cuts to determine which best replicates the progenitor samples' luminosity distributions. For this, we do not consider the SMC, whose observed lack of evolved stars have been discussed as an open issue \citep{Levesque_2006,Cheng_2024,massey_2000,heger_2003}.

\begin{figure*}
    \centerline{\includegraphics[width=0.95\textwidth]{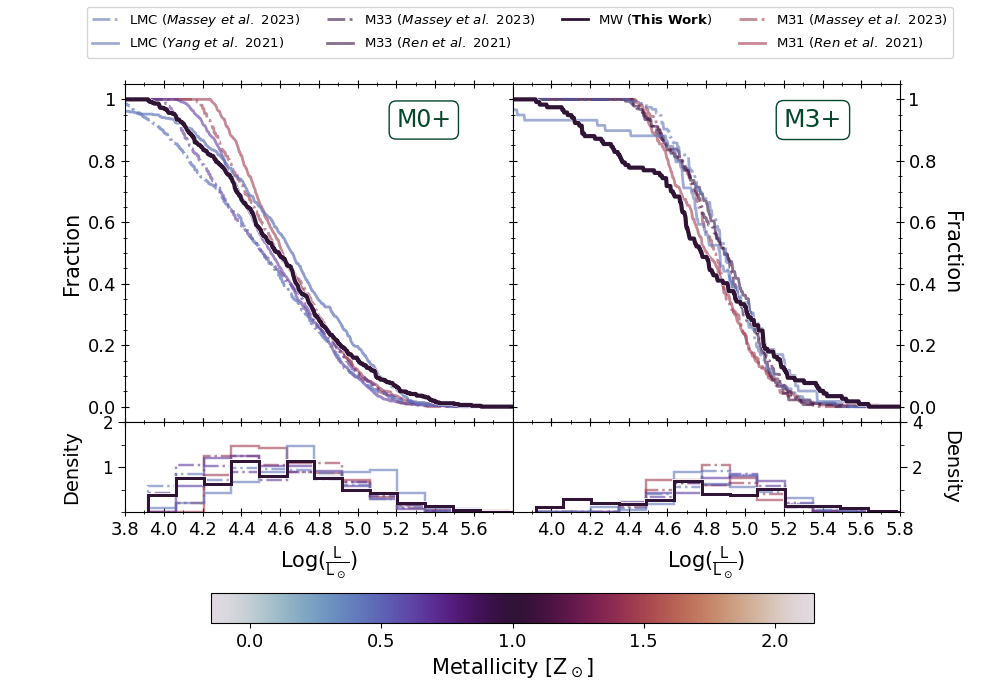}}
    \caption{Cumulative (top) and fractional (bottom) distributions of RSGs from MW, M31, M33, LMC, and SMC. The left shows all RSGs, while the right shows a sample limited to only objects with spectral type greater or equal to M3. The colormap shows the galaxy's average metallicity.}\label{fig: combined local group frac and cum comparison}
\end{figure*}

\begin{figure}
\centerline{\includegraphics[width=0.49\textwidth]{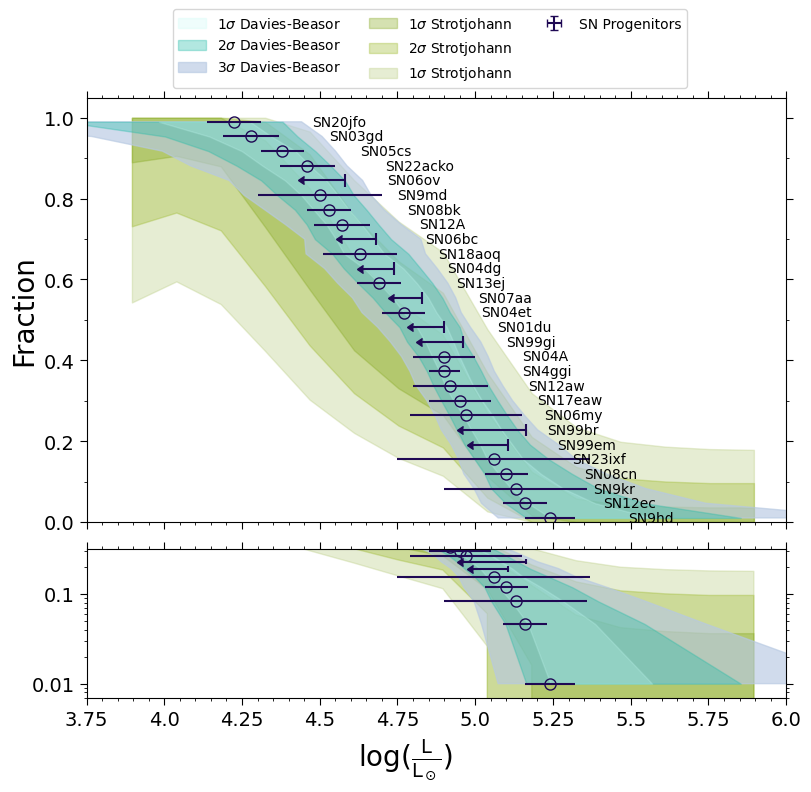}}
    \caption{The luminosity functions of SNe progenitors from Table \ref{A.2} using the ordering method. Included are both detections and upper limits, represented by black open circles and arrows, respectively. \cite{Strotjohann_2024}'s RSG progenitor luminosity function (shades of green) with its confidence regions due to statistical fluctuations and measurement uncertainties on the progenitor bolometric magnitudes. Along with \cite{davies_2020a_rsg_p_ii}, the luminosity function method applied to the updated progenitor list is shown in various shades of blue.}\label{fig:BCs against progenitors}
\end{figure}

\subsection{Luminosity distribution of RSGs} \label{sec: LUM of RSGs}

We produce CLDs and fractional luminosity distributions (FLD), as shown in Fig.~\ref{fig: combined local group frac and cum comparison}, for the MW and each galaxy in both the Ren and Massey lists. The comparison between the CLDs of these samples and our MW sample is broken into M0+ (left panel) and M3+ (right panel), and the color represents the metallicity (bottom color bar).

The MW sample has the most lenient cuts in order to compensate for the complications of extinction and uncertain distance, which are particularly apparent for M3+ side of Fig.~\ref{fig: combined local group frac and cum comparison} and the shallow slope in the lower luminosity range, but still retains information about metallicity effects and luminosity behavior. As expected based on sample criteria, the LMC CLD from \cite{yang_2021lmc} shows likely contamination from AGBs and foreground red giants (RGs), but M31 and M33 who are less likely to be affected by foreground RGs \citep{Massey_evans_2016} do not show the same excess of lower luminosity objects. As AGBs and RGs, even in the extreme, are restricted to the luminosity of mainly low luminosity, the difference in colour cuts shows little effect towards the high luminosity end. As a whole, the Ren and Massey samples agree within errors.  

We also find that, numerically, the maximum luminosities, listed in Table \ref{Tab:Tcr} columns 9-11, appear approximately independent of metallicity, settling at around $\rm log(L/L_\odot)=5.5$ for the Massey samples, and within uncertainties, also for the Ren sample, although it has a larger dispersion. However, examination of the lower end of the M0+ CLDs in Fig. \ref{fig: combined local group frac and cum comparison} indicates some metallicity dependence as the minimum luminosity increases with metallicity. This behavior is absent in the minimum luminosities of M3+ samples. While the dominant mechanism for SN progenitor mass loss is a subject of debate, this could suggest that for hotter low-mass RSGs, line-driven wind may be the primary driver of mass-loss, which transitions to other mechanisms at high masses and cooler temperatures.

\subsection{Luminosity distribution of SN pre-imaged RSGs} \label{sec: lum SN comparisons}
We build upon previous SN II-P and II-L pre-image lists originally based on \cite{smartt_2015} with later updates from \cite{beasor_2024} and \cite{Strotjohann_2024} (and references within), shown in Tables \ref{A.1} and \ref{A.2}. The distribution is shown in Fig.~\ref{fig:BCs against progenitors}, where the SNe are ordered based on luminosity and evenly distributed from 0 to 1, roughly reproducing the shape of a CLD. Note that while {SN2002hh} is included in our list of detected progenitors, we did not use it to constrain the minimum of our CLD's behavior since it only provides a limit.

\begin{figure*}
    \centerline{\includegraphics[width=0.95\textwidth]{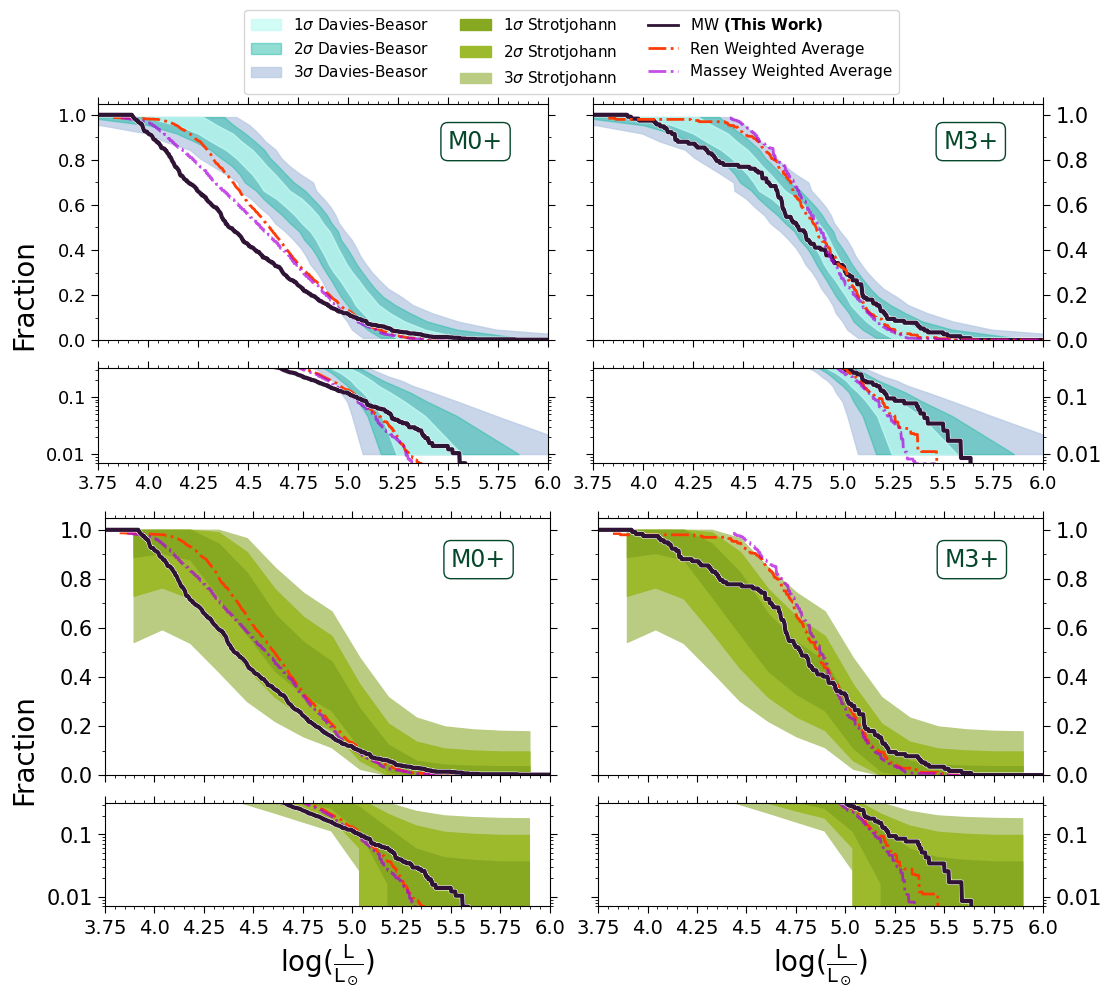}}
    \caption{SN pre-imaged progenitors, with corrections applied, compared to the RSGs CLD averaged based on metallicity. Top panels show the \cite{davies_2020a_rsg_p_ii} correction, while bottom panels show the \cite{Strotjohann_2024} correction. Left panels show $\geq$M0 observed RSGs (denoted M0+), while right panels show RSGs with the specific classification of $\geq$M3 observed stars (denoted M3+).}    
    \label{fig: DB and Strot versus weighted average}
\end{figure*}

Various corrections have been explored for SN pre-imaged progenitors. Firstly, \cite{davies_beasor_2018_bc} reworked the method of \cite{smartt_2009} and \cite{smartt_2015} by deriving updated bolometric corrections for RSGs and reevaluated the initial masses of all SNe progenitors \citep{davies_2020a_rsg_p_ii}. From those works, \cite{beasor_2024} determined the accuracy of the bolometric corrections for optical single-band derived luminosities as discussed in Sec.~\ref{sec:I lum}. With this result, they refined the previous progenitor luminosity function \citep{davies_2020a_rsg_p_ii} and for luminosity derived from F814W single-band observations add additional uncertainty. We follow the same Monte Carlo procedure; the resulting confidence intervals are shaded in blue in Figs.~\ref{fig:BCs against progenitors} and \ref{fig: DB and Strot versus weighted average}.

Secondly, \cite{Strotjohann_2024} incorporated measurement errors and sensitivity of the surveys when calculating a bias-corrected luminosity function for SN pre-imaged progenitors. They found a large increase in faint progenitors, e.g., at the highest end Mbol $\rm <-7$ mag reduced to $\rm 36\pm11\%$ from the previous estimate of $\rm 56\pm5\%$ of \cite{davies_2020a_rsg_p_ii}. We reproduce their results with the updated uncertainties of \cite{beasor_2024}, which are shown by shaded green regions in Figs.~\ref{fig:BCs against progenitors} and \ref{fig: DB and Strot versus weighted average}.

\subsection{Metallicity-averaged RSG distributions}\label{Sec:Z weighted averages}

The results found in Sec.~\ref{sec: LUM of RSGs} suggest portions of RSGs luminosity functions, particularly the lower luminosity end, are affected by metallicity. While the MW's metallicity is approximately that of the pre-imaged SN progenitor's average metallicity, it is not a complete sample. In order to determine how the  luminosity function of RSGs compares to that of the pre-imaged SN progenitors, we determine a second source which can be used to constrain how completeness effects the luminosity function. 

Using the estimated metallicties associated with each SN listed in Tab.~\ref{A.1} compiled from the literature as described in Sec.~\ref{pre-imaging data}, we break the sample into objects with metallicity which best match the average metallicity of M31, M33, LMC, and SMC. The sample sizes are then used as the weights to apply to the CLDs of M31, M33, LMC, and SMC to determine an average CLD that recovers the same metallicity profile as the pre-imaged SN progenitors and which is build of complete samples of RSGs. For both M0+ and M3+ samples, two weighted averages were built based on either the Ren or Massey lists. We retain the individual MW CLD since it shows the effect of AGB contamination, is at approximately solar metallicity, and provides a good point of reference.

The left side of Fig.~\ref{fig: DB and Strot versus weighted average} shows both \cite{beasor_2024} (top) and \cite{Strotjohann_2024}'s (bottom) corrections agianst the average CLD and MW for the M0+ samples. The right shows the same but for the M3+ sample. the result are discussed on Sec.~\ref{sec: results and dicussion}.

\section{Results and Discussion} \label{sec: results and dicussion}

\subsection{RSG spectral type at collapse} \label{sec:m3+}

For initial masses of $\rm \sim\!\!8\!-\!35 \,M_\odot$, the RSG phase is thought to be when most mass-loss occurs. The strong winds produce large amounts of dust that surround the RSG significantly altering the star and its appearance even to the point of potentially removing the entire H-rich envelope. As such, the mass loss rate governs the onward evolutionary path of the star and even what the supernova looks like. Due to the resulting changes, it is important to constrain the spectral types of SN progenitors as this affects the bolometric corrections and other assumptions used when deriving progenitors' luminosity. This also allows us to determine the correct sample of observed RSGs to compare to the pre-imaged SN progenitors and determine the significance of the RSG problem.

Figure \ref{fig: DB and Strot versus weighted average} compares various corrections to various subpopulations of RSGs. The \cite{beasor_2024} correction to SN pre-images are shown in the top panels while the \cite{Strotjohann_2024} bias corrections are shown in the bottom panels. The left-hand panels show these compared to the M0+ MW and metallicity weighted average CLDs from both the Ren and Massey lists (see Sec.~\ref{Sec:Z weighted averages} for details). In many places, the MW and weighted CLDs of M0+ RSG populations are beyond the 3$\sigma$ boundaries of the blue \cite{beasor_2024} correction (top-left) and with MW CLD at the $\rm 2-3\sigma$ and the weighted CLDs at the $\rm 1-2\sigma$ boundaries of the \cite{Strotjohann_2024} bias correction (bottom left). Based on Figs.~\ref{fig: nir_lum} and \ref{fig: I_lum}, we have shown that there is potentially more significant separation at the lower end as the weighted CLDs are likely over-estimated and the progenitors' corrections are likely underestimated. The use of NIR luminosity would shift the MW and weighted CLDs away from the $\rm 2-3\sigma$ or $\rm 1-2\sigma$, respectively, confidence intervals of both corrections at the luminous end, but this could be balanced out by the significant underestimation of uncertainties in for F814W derived luminosity. 

The right side of Fig.~\ref{fig: DB and Strot versus weighted average} shows comparisons to the MW and weighted CLDs of M3+ RSG population. These are within the 1$\sigma$ bounds of both \cite{beasor_2024} and \cite{Strotjohann_2024} corrections above $log(L/L_\odot) = 5.0$. At lower luminosity where the MW is known to have an excess of lower luminosity contaminates, the overlap with the \cite{Strotjohann_2024} correction is within the 1$\sigma$ and the \cite{beasor_2024} correction is between 2-3$\sigma$. As seen in Fig.~\ref{fig: nir_lum}, the weighted CLDs are likely overestimated at the lower luminosity end, suggesting the weighted CLDs should be closer to 1-2$\sigma$ of blue \cite{beasor_2024} correction and $\sim2\sigma$ green \cite{Strotjohann_2024}. This would also affect the higher luminosity end where the weighted CLDs and progenitor corrections are shown to agree within 1$\sigma$ at current estimates. Still, the NIR-caused shift of the weighted CLDs should be offset by the significant uncertainty increase in the corrections due to the F814W bolometric correction at later spectral types as seen in the higher luminosity end of Fig.~\ref{fig: I_lum}. For the entirety of the \cite{beasor_2024} correction, the M3+ CLDs is within 3$\sigma$ confidence, of which over half is within the 1$\sigma$ boundaries. 

While the \cite{Strotjohann_2024} bias correction does agree with previous results that show no missing high mass progenitors, it suggests $\rm \sim 10 \%$ of stars have luminosity low enough to imply ZAMS masses below the predicted limit for CCSNe ($\rm \sim6-8 M_\odot$). While we know the MW and to a lesser degree the weighted CLDs have some contamination from AGBS and foreground stars that are of lower masses, a list built off detected SNe later matched to archival imaging should not include estimates for objects that would be too small to produce CCSNe. An explanation for this is not addressed. Even when accounting for the added uncertainty in F814W-derived luminosity, the percent is still nonzero. Their results also produce significantly wider confidence intervals. For these reasons and the results seen in Fig.~\ref{fig: DB and Strot versus weighted average}, we favor the correction of \cite{beasor_2024}.

The weighted CLDs of M0+ stars do not produce the same shape and, for a good portion, disagree beyond the 3$\sigma$ boundaries of the \cite{davies_2020a_rsg_p_ii} correction. This suggests that M0+ RSGs have not evolved to have the same characteristics as the progenitors observed right before SNe. However, both of the weighted CLDs of M3+ stars match within 3$\sigma$ across the entire range, and greater than a half is within 1$\sigma$, suggesting that M3+ stars have a similar luminosity distribution to those of the pre-imaged detected SN progenitors.

We conclude that the M3+ subsamples are representative of the true luminosity function of Type II progenitors, with the shift being part of stellar evolution. Just prior to core collapse, the final stages of nuclear burning will cause the RSGs to brighten, moving upwards in the H-R diagram causing an overall shift to higher luminosity and lower mass RSGs spend fractionally less of their overall RSG lifetime at later spectral types creating the shift in luminosity function illustrated in Fig.~\ref{cum k0, m0, m3}. This supports previous suggestions in \cite{beasor_2024} (and reference within) that SN progenitors are exclusively RSG at the very end of the phase (M3 or later) on the basis that in star clusters with large numbers of RSGs, more evolved stars are observed to have later spectral types \citep{negueruela_2013,davies_beasor_2018_bc} with estimates of the typical spectral type at SN being M5-M7 and progenitors are shown clearly to be very red from multi-color pre-explosion photometry (SN2003gd, SN2004et SN2008bk, and SN2009md). 

Nevertheless, let us also consider the case where RSGs at the M0 stage could be the phase leading directly to SNe. If we take the \cite{davies_2020a_rsg_p_ii} correction, comparison to the weighted CLDs of M0+ stars suggests a missing number of faint pre-imaged SNe progenitors. However, the missing objects cover a luminosity range well below the predicted limit to produce CCSNe, suggesting this is not the case. If we take \cite{Strotjohann_2024} as the true correction, there would be a missing fraction of pre-imaged SN progenitors around the middle range of luminosities. This is unlikely the case as the authors do not currently know of a bias in any of the surveys or telescopes used to compile the sample of pre-imaged SN progenitors, which would favor only bright and faint objects and miss mean luminosity RSGs.  

\subsection{The Red Supergiant Problem}\label{sec:rsg problem}

The RSG problem describes the apparent missing population of Type II-P SN pre-images with ${\rm log}(L/L_\odot) > 5.24$ when compared with the maximum luminosity of observed field RSGs which reach to ${\rm log}(L/L_\odot) \sim 5.6$. 
We do not look into the RSG problem in terms of mass as previously done, as the additional uncertainties added when converting luminosity to $\rm M_{ZAMS}$ from stellar evolution models are large, and their effects are hard to trace. However, bolometric luminosity can be used to directly compare populations of observed RSG and confidence intervals of corrections to pre-imaged SN progenitors CLDs as shown in Fig.~\ref{fig: DB and Strot versus weighted average}.

For the comparison of the MW and weighted CLDs to pre-imaged SN progenitor corrections of \cite{davies_2020a_rsg_p_ii} in Fig.~\ref{fig: DB and Strot versus weighted average}, the MW and weighted CLDs of M3+ RSG population are within the 1$\sigma$ bounds of both corrections above ${\rm log}(L/L_\odot) = 5.0$. At the upper luminosity end, Figs.~\ref{fig: nir_lum} and \ref{fig: I_lum} show a significant increase in uncertainty not accounted for in the current estimates of both the RSG luminosities and the pre-imaged SN progenitors, but this is unlikely to shift the weighted CLDs out of the boundaries of the 1$\sigma$ confidence intervals from \cite{davies_2020a_rsg_p_ii} method. The agreement, even when accounting for derivation issues, confirms that there are no missing high-luminosity RSGs.

\subsection{The Lower Luminosity End}\label{sec:low lum problem}

As more faint and dusty RSGs have become observable in large-scale surveys, samples of RSGs have become more complete to lower luminosity limits. This has led to a suggestion of an apparent disagreement at the lower luminosity end \citep{beasor_2024}. While the initial mass must be above $\rm \sim8M_\odot$ for a single star to undergo CC \citep{heger_2003}, some pre-SN imaged progenitor corrections imply the boundary is lower. But our results show the M3+ MW CLD below or at the 2$\sigma$ boundary of the \cite{beasor_2024} correction in this region, indicating that the lowest estimated progenitor luminosity is still greater than that of a sample of RSGs with extreme AGBs whose masses are $\rm \sim6-8M_\odot$ contamination and it is therefore more likely the luminosity is unreliable then there begin a contradiction to stellar theory.

While not in complete agreement, the weighted averages and CLDs of M3+ sample are within the 3$\sigma$ confidence interval above the mean of the \cite{davies_2020a_rsg_p_ii} correction. When considering the effect of NIR single-band luminosity derivation, this gap shrinks. The lower luminosity end has the NIR overestimated, bringing the weighted CLDs down, and $\rm BC_I$ for M3+ RSGs increases the uncertainty in the luminosity and widens the boundaries, especially as the three lowest pre-imaged SN progenitors's luminosity comes from F814W. Between the weighted CLDs and the M3+ MW CLD below or at the 2$\sigma$ boundary of the \cite{beasor_2024}, we find no evidence that there is a missing low-mass SN progenitor problem.

\section{Conclusion} \label{sec: conclude}

We have performed a re-assessment of the RSG problem using well-mapped sample of RSGs from the Local Group which covers a range of metallicities. Our main results are summarized below. 

\begin{itemize}

\item When considering RSG with spectral types $\geq$M0, we found a metallicity dependence of the minimum luminosity, suggesting line-driven wind may be the dominant mechanism for RSG mass-loss at lower masses and hotter temperatures but transitions to other mechanisms at high masses and cooler temperatures.

\item RSGs who display luminosity behaviors evident of impending CC are constrain to spectral types at or later then M3.

\item We have found no evidence of high luminosity SN progenitors missing among SN pre-images. We find that the uncertainties on the single-band derived luminosity of pre-imaged SN progenitors and complete Local Group samples of RSGs are too large to meaningfully infer population differences.

\item While some disagreement is found at low luminosities between Local Group RSGs and pre-imaged SN progenitors, this can be explained by the inaccuracy and uncertainty of bolometric corrections in the optical and NIR.

\end{itemize}

Our knowledge of evolved massive stars still has large gaps, and much research needs to be done to better understand the luminosities of RSGs. For example, we cannot account for the additional uncertainties from outstanding issues such as binary interactions, mass loss rates, and black hole mass thresholds. Nevertheless, large surveys are making it possible to observe more complete populations of RSGs, and the upcoming era of NIR large-scale surveys means that questions regarding the dust production of these evolved stars can be informed by better observations. 

In particular, future efforts to determine bolometric corrections that are more attuned to evolved massive stars for bands in which progenitors have been and are likely to be imaged in will be particularly important.

\vspace{12mm}

S.~Healy is supported by NSF Grant No.~PHY-2209420. The work of S.~Horiuchi is supported by the U.S.~Department of Energy Office of Science under award number DE-SC0020262, NSF Grant No.~AST1908960 and No.~PHY-2209420, and JSPS KAKENHI Grant Number JP22K03630 and JP23H04899. This work was supported by World Premier International Research Center Initiative (WPI Initiative), MEXT, Japan.


\bibliography{bib}{}
\bibliographystyle{aasjournal}

%






\appendix

\section{Supernova Tables}

We provide details of our SN pre-imaged progenitor sample in Tables \ref{A.1} and \ref{A.2}. 

\begin{deluxetable*}{lccccc}[t]
\tabletypesize{\scriptsize}
\tablewidth{0pt} 
\tablenum{A.1}
\tablecaption{SN Pre-imaged Progenitor Sample Parameters \label{A.1} }
\tablehead{SN &  Sptype & $\rm T_{\rm eff}$ & Host Galaxy & 12 + log(O/H) & Notes\\
&  & [K] & & &}
    \colnumbers
    \startdata
    1999an & assumed M-type & 3300-3790 &IC 755	&	8.3 \\
    1999br &  assumed M & 3300-3790 & NGC 4900	&	8.4  \\
    1999em & K2–M4 & 3400-4015 & NGC 1637 & 8.6 \\
    1999gi &  M & 3300-3790 & NGC 3184 & 8.6\\
    2001du &  K2–M4 & 3400-4015 & NGC 1365 & 8.5 \\
    2002hh &  assume K0-M5 & 3300-4185 & NGC 6946 & 8.5\\
    2003gd & M3, K5–M3 & 3605.0 & NGC 628 & 8.4	& HII regions in M51 display near solar metallicity \\
    2004A & M-type, G5–M5 & 3300-4185 & NGC 6207 & 8.3\\
    2004dg & assumed RSG & 3300-4185 & NGC 5806 & 8.5 \\
    2004et &  $\rm >$M4 & 3300-3450 & NGC 6946 & 8.3\\
    2005cs &  K5-M4 & 3600.0 & NGC 5194 & 8.7\\
    2006bc & assumed RSG & 3300-4185 & NGC 2397 & 8.5\\
    2006my & assumed RSG & 3300-4185 & NGC 4651 & 8.7\\
    2006ov & assumed RSG & 3300-4185 & NGC 4303 & 8.9\\
    2007aa &  M-type & 3300-3790 & NGC 4030 & 8.4\\	
    2008bk &  $\geq$M4, Late-type M & 3300-3535 & NGC 7793 & 8.4 & estimated between SMC and LMC \\
    2008cn &  YSG  & 5200.0 & NGC 4603 & $8.76\pm0.24$\\
    2009hd & early K & 4500.0 & NGC 3627	& $8.43\pm0.05$\\
    2009md &  M4 & 3300.0 & NGC 3389 & $8.96\pm0.04$ &	estimate based on mag absolute and radial z gradient \\
    2009kr & RSG or YSG & 2500-5200 & NGC 1832 & $8.55\pm0.16$\\
    2012A &  $\sim$ late-type RSG & 3300-3535 & NGC 3239 & $8.11\pm0.09/8.04\pm0.01$\\
    2012aw &  RSG & 3300-4185 & M95 (NGC 3351) & $8.8\pm0.1$ \\
    2012ec &  $>$ K0 & $4500^*$ & NGC 1084 & $8.93\pm0.1$ \\
    2013ej & M2 & 3800.0 & M74 & 8.73 \\
    2017eaw & RSG & 2500–3300 & NGC 6946 & 8.51\\
    2018aoq &  M-type RSG & 3500.0 & NGC 4151 & & no reliable estimation\\
    2020jfo & RSG &  3300–3500 & M61 & & M31-like (1.5-2$\rm z_\odot$)\\
    2022acko &  Likely RSG &  3300-4185 & NGC 1300 & &	approximately solar metallicity, perhaps slightly higher \\
    2023ixf &  RSG & 3920.0 & M101 & & subsolar to supersolar \\
    2024ggi & late-type RSG & $3290^{+19}_{-27}$ & NGC 3621 & $8.23\pm0.05$ & \\
    \enddata
\tablecomments{Summary of Progenitor spectral type (Sptype), effective temperature ($T_{\rm eff}$), host galaxy, and metallicity (O/H) compiled from the literature}
\footnotesize{$^a$ $T_{\rm eff}$ placed are rough estimates used when placing on H-R diagram.

$^b$ Assumed RSGs means that estimates for values are derived based on the assumption the progenitor is a RSG others come from best fits to SEDs and spectral analysis. 

}
\end{deluxetable*}

\begin{deluxetable*}{lcccccccc}
\tabletypesize{\scriptsize}
\tablewidth{0pt} 
\tablenum{A.2}
\tablecaption{SN Pre-imaged Progenitor Sample Parameters \label{A.2} }
\tablehead{SN &  DM & $\rm \lambda$ & $\rm A_{\lambda}$ & $\rm BC_\lambda$ & $\rm m_\lambda$ & $\rm M_{bol}$  & $\rm Log(\frac{L}{L_\odot})$ & lim\_l\\
& [mag] & &   & [mag] & [mag] & [mag]  & &}
    \colnumbers
    \startdata
    1999an & $31.34 \pm0.08$ & F606W & $0.28\pm0.13$ & $-1.83\pm0.38$ & $>24.7^{0.1}$ &  $>-8.7^{0.41}$ &   &   $5.38^{0.164}$ \\
    1999br &   $30.75\pm0.18$ & F606W & $0.04\pm0.04$ & $-1.83\pm0.38$ & $>24.9^{0.1}$ & $>-7.71^{0.43}$ &  &  $4.98^{0.17}$ \\
    1999em &  $30.34\pm0.09$ & Ic & $0.16\pm0.08$ & $-0.32\pm0.15$ & $>23.0^{0.1}$ & $>-7.79^{0.21}$ &   & $5.01^{0.08}$ \\
    1999gi &  $30.0\pm0.08$ & F606W & $0.45\pm0.11$ & $-1.83\pm0.38$ & $>24.9^{0.1}$ & $>-7.3^{0.41}$ & & $4.85^{0.1}$ \\
    2001du &  $31.31\pm0.07$ & F814W & $0.26\pm0.14$ & $0.0\pm0.15$  & $>24.25^{0.1}$ & $>-7.32^{0.24}$ &  & $4.82^{0.07}$ \\
    2002hh &   $29.43\pm0.12$ & i & $2.6\pm0.1$ &
    $-0.49\pm0.15$ & $>22.8^{0.1}$ & $>-9.23^{0.23}$ &  & $5.59^{0.9}$ \\
    2003gd & $29.84\pm0.19$ & F814W & $0.23\pm0.07$ & $0.1\pm0.15$ & $24.0\pm0.04$ & $-5.97\pm0.26$ & $4.28\pm0.09$ & $3.33^{0.11}$ \\
    2004A & $31.54\pm0.17$ & F814W & $0.34\pm0.1$ & $0.0\pm0.15$ & $24.36\pm0.12$ & $-7.52\pm0.28$ & $4.9\pm0.1$ & $4.42^{0.11}$ \\
    2004dg & $31.51\pm0.13$ & F814W & $0.4\pm0.05$ & $0.0\pm0.15$ & $>25^{0.1}$ & $>-6.91^{0.23}$ & &$4.66^{0.07}$\\
    2004et &  $29.43\pm0.12$ & Ij & $0.64\pm0.05$ & $0.25\pm0.15$ & $22.06\pm0.12$ & $-7.78\pm0.23$ & $4.77\pm0.07$& $4.53^{0.09}$ \\
    2005cs &  $29.62\pm0.12$ & F814W & $0.27\pm0.03$ & $0.05\pm0.15$ & $23.62\pm0.07$ & $-6.27\pm0.21$ & $4.38\pm0.07$ & $3.7^{0.09}$ \\
    2006bc &  $30.84\pm0.18$ & F814W & $0.34\pm0.0$ & $0.0\pm0.15$ &$>24.45^{0.1}$ & $>-6.75^{0.26}$ & &$4.59^{0.08}$ \\
    2006my &  $31.74\pm0.12$ & F814W & $0.81\pm0.42$ &  $0.05\pm0.15$& $24.86\pm0.13$& $-7.7\pm0.48$& $4.97\pm0.8$ & $4.53^{0.19}$\\
    2006ov &  $30.5\pm0.19$ & F814W & $0.18\pm0.04$ &  $0.05\pm0.15$& $>24.2^{0.1}$& $>-6.43^{0.26}$& & $4.47^{0.1}$\\
    2007aa &  $31.55\pm0.13$ & F814W & $0.05\pm0.01$ & $0.05\pm0.15$& $>24.44^{0.1}$ & $>-7.17^{0.23}$& & $4.76^{0.06}$\\
    2008bk &  $27.96\pm0.13$ & K & $0.03\pm0.01$ &$3\pm0.18$ &$18.38\pm0.03$ & $-6.6\pm0.22$& $4.53\pm0.07$ & $3.46^{0.10}$\\
    2008cn &  $32.61\pm0.1$ & F814W & $0.54\pm0.06$ &$0.05\pm0.15$& $25.13\pm0.09$ & $-8.02\pm0.21$ & $5.1\pm0.07$ & $4.5^{0.09}$\\
    2009hd &  $29.86\pm0.08$ & F814W & $2.04\pm0.08$& $0.05\pm0.15$& $23.54\pm0.14$ & $-8.36\pm0.23$ & $5.24\pm0.08$ & $4.83^{0.09}$\\
    2009md &   $31.64\pm0.21$ & F814W & $0.17\pm0.17$ & $0.05\pm0.15$ & $24.87\pm0.11$& $-6.94\pm0.33$& $4.5\pm0.2$ & $4.14^{0.13}$\\
    2009kr &  $32.09\pm0.5$ & F555W & $0.58 \pm0.05$ & $-0.36\pm0.2$ & $24.71\pm0.23$ & $-8.32\pm0.59$&  $5.13\pm0.23$ & $5.10^{0.22}$ \\
    2012A &   $29.96\pm0.15$ & K & $0.01\pm0.0$ & $3\pm0.18$ & $20.29\pm0.13$ & $-6.68\pm0.27$ & $4.57\pm0.09$ & $4.12^{0.1}$\\
    2012aw &   $29.96\pm0.2$ & K & $0.16\pm0.02$ &$3\pm0.18$ & $19.56\pm0.29$ & $-7.56\pm0.34$ & $4.92\pm0.12$ & $4.82^{0.08}$\\
    2012ec &   $31.19\pm0.1$ & F814W & $0.36\pm0.1$ &  $0.0\pm0.15$ &$23.39\pm0.08$ & $-8.16\pm0.22$ & $5.16\pm0.07$& $4.5^{0.09}$\\
    2013ej &  $29.8\pm0.11$ & F814W & $0.24\pm0.03$& $0.4\pm0.2$& $22.65\pm0.05$ & $-6.99\pm0.24$ & $4.69\pm0.07$ & $3.83^{0.1}$\\
    2017eaw &  $29.44\pm0.12$ & F606W & $0.74\pm0.09$ & $-1.83\pm0.38$ & $26.4\pm0.05$ & $-5.53\pm0.41$ & $4.95\pm0.1$ & $3.25^{0.17}$\\
    2018aoq &   $31.3\pm0.12$ & F814W & $0.05\pm0.07$ & $0.0\pm0.15$ & $23.91\pm0.02$ & $-7.44\pm0.2$ & $4.63\pm0.12$ & $3.71^{0.09}$\\
    2020jfo &  $30.81\pm0.12$ & F814W & $0.03\pm0.07$ & $0\pm0.15$& $25.02\pm0.07$ & $-5.82\pm0.22$ & $4.22\pm0.09$ & $3.51^{0.09}$\\
    2022acko &   $31.74\pm0.12$ & F814W & $0.28\pm0.07$ & $0\pm0.15$& $25.61\pm0.09$ & $-6.41\pm0.22$& $4.46\pm0.09$ & $3.86^{0.09}$ \\
    2023ixf &   $29.18\pm0.05$ & F814W & $0.07\pm0.07$ & $-2.25\pm0.1$ & $24.44\pm0.06$ & $-7.09\pm0.14$& $5.06\pm0.31$ & $3.95^{0.06}$\\
    2024ggi &  $29.14\pm0.06$ & F814W & $0.32\pm0.0$ & & 23.25& $-6.21\pm0.08$ & $4.9\pm0.05$ & $3.58^{0.05}$\\
    \enddata
\tablecomments{Summary of Progenitor Characteristics compiled from the literature. Showing the SN's name (1), distance modulus (2), bands used to derived luminosity (3), associated extinction (4), associated bolometric correction (5), observed magnitude (6), bolometric magnitude (7), bolometric luminosity (8), and limiting magnitude (9) as determined in Sec \ref{pre-imaging data}. \\
\textbf{References}: \cite{beasor_2024} and references within}

\footnotesize{$^a$ $T_{\rm eff}$ placed are rough estimates used when placing on H-R diagram.

$^b$ Assumed RSGs means that estimates for values are derived based on the assumption the progenitor is a RSG others come from best fits to SEDs and spectral analysis.}
\end{deluxetable*}

\end{document}